\documentclass[pdflatex,sn-aps]{sn-jnl}

\usepackage{graphicx}%
\usepackage{multirow}%
\usepackage{amsmath,amssymb,amsfonts}%
\usepackage{amsthm}%
\usepackage{mathrsfs}%
\usepackage[title]{appendix}%
\usepackage{xcolor}%
\usepackage{textcomp}%
\usepackage{manyfoot}%
\usepackage{booktabs}%
\usepackage{algorithm}%
\usepackage{algorithmicx}%
\usepackage{algpseudocode}%
\usepackage{listings}%

\theoremstyle{thmstyleone}%
\theoremstyle{thmstyletwo}%

\theoremstyle{thmstylethree}%

\begin{document}

\title[Geometric Construction of Dynamically Corrected Quantum Gates]
{Geometric Construction of Dynamically Corrected Quantum Gates}


\author*[1]{\fnm{Shingo} \sur{Kukita}}\email{kukita@nda.ac.jp}
\equalcont{These authors contributed equally to this work.}

\author*[2]{\fnm{Yasushi} \sur{Kondo}}\email{kondou.yasushi.m50@kyoto-u.jp}
\equalcont{These authors contributed equally to this work.}

\affil[1]{\orgdiv{Depart. of Computer Science}, 
	\orgname{National Defence Academy of Japan}, 
	\orgaddress{
		 \city{Yokosuka}, \postcode{239-8686}, 
		 \country{Japan}}}

\affil[2]{\orgdiv{Physics Depart.}, \orgname{Kindai University}, 
	 \orgaddress{
	 	\city{Higashi-Osaka}, \postcode{577-8502}, 
	 	\country{Japan}}}


\abstract{
The foundation of quantum technologies lies in the precise control of quantum systems. It is crucial to implement dynamically corrected quantum gates (DCQG), which compensate for individual quantum gate errors to make them more resilient to errors alongside quantum error correction. Off-resonance error (ORE), which originates from fluctuation and mis-calibration of resonance frequencies of qubits, is one of the most critical error types to be compensated. There have been many studies on constructing DCQGs robust against ORE up to its first order.Explicit construction of second-order robust DCQGs against ORE has been discussed less. Recently, the geometric meaning of the second-order robustness against ORE was uncovered. From this implication, we propose a geometric construction of second-order DCQGs against ORE using a first-order DCQG as a seed.
}




\maketitle

\section{Introduction}\label{Intro}

In recent years, application of quantum technologies such as quantum computing \cite{Nielsen2000,bennett2000quantum, nakahara2008quantum}, metrology \cite{helstrom1976quantum,caves1981quantum,holevo2011probabilistic}, and communications \cite{ekert1991quantum,gisin2007quantum,chen2021integrated} have been attracting significant attention. To realize practical quantum technologies, a hierarchical architecture of quantum information processing was proposed in Ref.~\cite{PhysRevX.2.031007}. In this structure, we gradually reinterpret physical qubits into logical qubits. The lowest layer of the structure is the physical layer, which contains many physical qubits. We need to employ open-loop error mitigation techniques, such as dynamical decoupling and dynamically corrected quantum gates (DCQGs), in this layer to suppress physical errors. Suppressing physical errors makes it effective to implement feedback-based quantum error correction \cite{lidar2013quantum}, which enables us to realize flawless logical qubits. The foundation of quantum technologies lies in the precise control of quantum systems, combined with open-loop error mitigation.

The Hamiltonian of qubits in real quantum systems slowly drifts, which causes undesired systematic error in operations. This slow drift is difficult to eliminate only by hardware calibration. One of the software error cancellation techniques is DCQGs, which are also called composite quantum gates or composite pulses in some literature \cite{counsell1985analytical,levitt1986composite,claridge2016high,Levitt2008}. One obtains a DCQG by replacing an operation with a (discrete or continuous) sequence of other operations so that systematic error is canceled at the end of the operation. Many DCQG sequences that are robust against some specific errors have been found \cite{brown2004arbitrarily,wimperis1994broadband,Cummins_2000,PhysRevA.67.042308}. The efficacy of DCQGs has been evaluated broadly, e.g., in nuclear magnetic resonance \cite{SR_Comp_Pulse_2020}, trapped ions \cite{PhysRevA.77.052334,PhysRevX.7.041061,Ball_2021,PhysRevA.106.062617}, superconducting qubits \cite{collin2004nmr,Ball_2021,PhysRevApplied.18.034062}, quantum dots \cite{PhysRevX.8.021058}, optical clocks \cite{Zanon-Willette_2018} and nitrogen vacancy centers \cite{PhysRevA.80.032303}.

DCQGs have been studied from geometrical aspects \cite{https://doi.org/10.1002/qua.24941,PhysRevA.80.024302,doi:10.1143/JPSJ.80.054002,Kukita:2022aa}. Pulse length error (PLE), which arises from imperfect calibration of the strength of a control field, was found to be related to the Aharonov-Anandan geometric phase \cite{doi:10.1143/JPSJ.80.054002,Kukita:2022aa}. More precisely, the first-order term of PLE in the whole operation is described by the geometric phase. Therefore, one can obtain a DCQG robust against PLE by designing a quantum gate so that this phase disappears: This is an example of ``geometric design" of DCQGs.  Another essential type of systematic error is off-resonance error (ORE), which originates from imperfect calibration of the resonance frequency of the target qubit. This error is also related to geometric quantity: The first-order term of ORE is associated with the geometry of the path representing the dynamics of a quantum state in the Bloch sphere \cite{Kukita:2022aa}. It was found that the second-order term of ORE also has a geometric interpretation \cite{Zeng_2018}.

In this paper, we propose a method to promote a first-order ORE-robust DCQG to a second-order one using the geometrical implication. We demonstrate the technique to ``promote" a famous ORE-robust DCQG, called CORPSE \cite{Cummins_2000,PhysRevA.67.042308}. CORPSE is a family of simple DCQGs. The shortest one among this family, called Short-CORPSE, is useful owing to its short operation time. The time optimality of Short-CORPSE has been discussed from some points of view \cite{PhysRevA.106.042613,Kukita_2024}. We add a simple operation into the Short-CORPSE sequence and obtain a second-order ORE-robust DCQG. Our promotion method applies to any first-order ORE-robust DCQGs, and therefore will be practical. 

This paper is organized as follows. Section~\ref{Theory} introduces a fundamental theory of DCQGs and the geometrical interpretation of second-order ORE robustness. Using this interpretation, we construct a second-order DCQG by modifying short-CORPSE in Sec~\ref{sec:ORE2nd}. Section~\ref{conc} is devoted to a summary of this work.
 
\section{ORE-robust DCQG}
\label{Theory}

To construct DCQGs, the effects of these errors are expanded in a series, and the gates are designed so that the coefficients of these terms become zero. Consider a Hamiltonian that implements a desired operation in the absence of errors. Let the time evolution operator under this Hamiltonian be denoted as $U(t)$, where $t$ moves in $t \in [0, T]$. $U(t)$ becomes $U_\delta(t)$ under the influence of errors, where $\delta$ represents the error in control parameters. In the interaction picture without errors, this time evolution operator can be written as $U^{\rm I} (t) = 1$, representing the identity operator. Note that $*^{\rm I}$ denotes ``interaction picture''. On the other hand, when errors are present, the operator can be expanded as
\begin{align}	
	U^{\rm I}_\delta (t) = 1+ \sum_i U_\delta^{(i)} (t) \delta^i,
\end{align}
where $U_\delta^{(i)} (0) = 0 $ but that generally $U_\delta^{(i)} (T) \ne 0$. We define a DCQG as follows. 
\begin{quote}	
	A $n$th-order robust DCQG satisfies $U_\delta^{(i)} (T) = 0$ for all $i\le n$.
\end{quote}
Note that $U_\delta^{(i)} (t)$ is not necessarily $0$ when $t \in (0, T)$. 

We focus hereinafter on 1-qubit DCQGs that are robust against ORE and closely follow Ref.~\cite{PhysRevA.98.012301,Zeng_2018}. As mentioned above, ORE originates from imperfect calibration in the energy scale of the qubit. When the $x$-axis is taken to be the quantization axis while the $y,z$-axes are taken to be the control axes, the Hamiltonian is given by
\begin{align}
	H(t) &=\Omega(t) \cos\phi\frac{\sigma_z}{2}+\Omega(t) \sin\phi\frac{\sigma_y}{2} + \delta \frac{\sigma_x}{2}.
	\label{eq:hamiltonian}
\end{align}
where $\Omega(t)$ is the signed strength of the driving field and $\phi$ is the phase. Note that the $z$-axis is often taken as the quantization axis, and the other two axes are the control axis. However, this is just a difference of terminology; they are always interchangeable.

It is well-known that ORE can be mitigated by one-axis control \cite{Cummins_2000,PhysRevA.67.042308}. Hence, we hereinafter consider the following type of control sequences:
\begin{align}
	H(t)&=\Omega(t) \frac{\sigma_z}{2}+ \delta \frac{\sigma_x}{2}.
\end{align}
We expand $U_\delta(t)$ up to  the $\delta^2$ terms as
\begin{align}
	U_\delta(t)&= \begin{pmatrix}
		u_1(t) & -u_2^*(t) \\ u_2(t) & u_1^*(t)
	\end{pmatrix},
	\label{eq:parameterize} 
\end{align}
where 
\begin{align}
	u_1(t) &= e^{-i\phi(t)/2 } \left( g_0(t) +f_1(t)\delta/2 +g_2(t)\left(\delta/2\right)^{2} + O(\delta^3) \right),
	\nonumber \\
	u_2(t) &= -i e^{i\phi(t)/2 }\left(f^{*}_{0}(t)+g_1^*(t)\delta/2 + f_2^*(t) \left(\delta/2\right)^{2} + O(\delta^3)  \right),
	\nonumber \\
	\phi(t) &=  \int_0^t \Omega(\tau) d\tau.
\end{align}
The Schr\"odinger equation of the time development operator $U_\delta(t)$ is 
\begin{align}
	i \dot{U}_\delta(t) &= H(t) U_\delta(t).
\end{align}
By comparing the coefficients of the $\delta^n$ terms on both sides of the Schr\"odinger equation, we obtain the following equations for a general driving field $\Omega(t)$.
\begin{align*}
\dot{f}_0(t) &=0,&\dot{f}_1(t) &=-e^{i \phi(t)}f^{*}_{0}(t), & \dot{f}_2(t) &=e^{i \phi(t)}f^{*}_{1}(t),
\nonumber \\
\dot{g}_0(t) &=0,&\dot{g}_1(t) &=e^{i \phi(t)}g^{*}_{0}(t), & \dot{g}_2(t) &=-e^{i \phi(t)}g^{*}_{1}(t).
\end{align*}
The initial condition $g_{0}(0)=1~(f_{0}(0)=0)$ and the equation $\dot{g}_0(t) =0~(\dot{f}_0(t) =0)$ lead to $g_{0}(t)=1~(f_{0}(t)=0)$. $f_1(0) =0$ and $\dot{f}_1(t) =0$ leads $f_1(t) =0$. Similarly, $f_2(t) =0$ is obtained. We hereafter employ the following relations. 
\begin{align}
	\label{rec_rel}
	\dot{g}_1(t)=  e^{ i\phi(t) }g^*_{0}(t),  & \,\,\, \dot{g}_2(t)=  -e^{ i\phi(t) }g^*_{1}(t).
\end{align}

The unitarity $U_\delta(t)U^{\dagger}_\delta(t)={\mathbb I}_2$, where ${\mathbb I}_2$ is the 2-dimensional identity matrix, can be explicitly calculated as 
\begin{align*}   
	& |u_1|^2 + |u_2|^2 \nonumber \\
	&= 
	\left(g_0(t) +g_2(t)\delta^2 +O(\delta^4) \right) \left(g_0(t) +g_2^*(t)\delta^2+ O(\delta^4) \right)+ g_1^*(t)g_1(t) \delta^2 + O(\delta^4)=1,
\end{align*}
which implies
\begin{align}	
	|g_1(t)|^2 = -(g_2(t)+g_2^*(t)). 
\end{align}
From this equation, one find that $\Re(g_2(T)) = 0 $ if $g_1(T) =0$. 

We parameterize $g_1(t)$ as 
\begin{align}	
	\label{eq9}	
	g_1(t) &= x(t) + i y(t),~x(t),y(t)\in {\mathbb R},
\end{align}
and consider the {\it error trajectory} $(x(t), y(t))$ in a two-dimensional space. The first-order error vanishing condition $g_{1}(T)=0$ implies that the trajectory $(x(t),y(t))$ is closed. Furthermore, this error trajectory has information not only on the first-order error but also on the pulse shape and the second-order error, as shown below.

From Eq.~\eqref{rec_rel}, one obtains
\begin{align}
	\label{rec_rel_1}
	\dot{g}_1(t) &= \dot{x}(t) + i \dot{y}(t)= e^{i\phi(t)} g_0^*(t)= e^{i\phi(t)},
\end{align}
which leads to
\begin{align}
	\dot{x}^2 + \dot{y}^2 = \dot{g}_1(t) \dot{g}^{* }_1(t)&= 1. 
	\label{xycond}
\end{align}
By differentiating both sides of Eq.~\eqref{rec_rel_1}, we obtain
\begin{align*}
	\ddot{g}_{1}(t)=\ddot{x}+i \ddot{y}(t) &= -i \Omega(t) e^{i\phi(t)}
	=-i  (\dot{x} + i \dot{y} )\Omega(t). 
\end{align*}
From this equation, $\Omega(t)$ is rewritten as
\begin{align}
	\Omega(t) &= \frac{\ddot{x}+i\ddot{y}}{-i(\dot{x}+i\dot{y})}
	= i \frac{(\ddot{x}+i\ddot{y})(\dot{x}-i\dot{y})}{\dot{x}^{2}+\dot{y}^{2}}
	=i \left((\ddot{x}\dot{x}+\ddot{y}\dot{y})-i\ddot{x}\dot{y}+i\ddot{y}\dot{x}\right)
	=  \ddot{x} \dot{y}-\dot{x}\ddot{y} 
	\nonumber\\
	&= \frac{  \ddot{x} \dot{y}-\dot{x}\ddot{y}}{(\dot{x}^2+ \dot{y}^2)^{3/2}},
\end{align}
where we use Eq.~\eqref{xycond}. Thus, $\Omega(t)$ is regarded as the instantaneous signed curvature of the trajectory $(x(t), y(t))$; the trajectory moves along a circle with radius $1/|\Omega(t)|$ clockwise if $\Omega(t)>0$ and counter-clockwise if $\Omega(t)<0$ at $t$. Note that $(\dot{x}(0),\dot{y}(0))=(1,0)$ for any control sequence. The trajectory starts from the origin with the velocity directed to the $x$-axis. Its perimeter represents the operation time $T$:
\begin{align}
T&=\int^{T}_{0}dt = \int^{T}_{0} \left(\dot{x}^2 + \dot{y}^2\right) dt.
\end{align}

Although the above statements are valid when the pulse shape $\Omega(t)$ changes continuously, we hereinafter consider piece-wise constant pulse shapes; i.e., $\Omega(t)=\Omega_{1} (t_{0}:=0\leq t < t_{1}),~\Omega_{2} (t_{1} \leq t < t_{2}) \cdots$, where $\Omega_{i}$ are constant because we later discuss improvement of  the Short-CORPSE \cite{Cummins_2000,PhysRevA.67.042308}. In this case, the trajectory at the time $t~(t_{i-1}\leq t <t_{i})$ moves on an arc with (signed) radius $R_{i}=1/\Omega_{i}$. The (zeroth-order) rotation angle $\theta_{i}$ in the Bloch sphere representation during this control is
\begin{align}
	\theta_{i}&=\int^{t_{i}}_{t_{i-1}}\Omega_{i}d t =\Omega_{i}(t_{i}-t_{i-1})=\frac{t_{i}-t_{i-1}}{R_{i}},
\end{align}
or equivalently, the center angle of the arc. Thus, the trajectory of $g_{1}(t)$ has the information on the zeroth-order ideal development.

Then we explain how the information on the second-order error and the corresponding robustness is encoded in the trajectory of $g_{1}(t)$. When $g_{1}(T)=0$, it is sufficient to consider $\Im(g_2(T))$ as discussed above. From Eq.~\eqref{rec_rel}, we obtain
\begin{align}
	\Im(g_2(T))  &= -\Im\left( \int_0^T e^{-i \phi(\tau )} g_{1}^*(\tau)d\tau \right) 
	= -\Im\left( \int_0^T (\dot{x} +i\dot{y} ) (x -i y)d\tau \right) 
	\nonumber \\
	&= -\Im\left( \int_0^T (\dot{x} x -i \dot{x} y + i \dot{y} x - \dot{y} y) d\tau \right) 
	\nonumber \\
	&= -\int_0^T (  x \dot{y}  -\dot{x} y  ) d\tau=-2 S.
\end{align}
where $S$ is the (signed) area enclosed by the trajectory according to Green's theorem.
Hence, when the first-order robustness is satisfied, the condition for the second-order robustness is that the zero net area.

\section{Turning Short-CORPSE to second-order DCQG}
\label{sec:ORE2nd}

We apply the above observation to provide the second-order robustness for Short-CORPSE \cite{Cummins_2000,PhysRevA.67.042308}, which is one of the most well-known first-order DCQG robust against ORE. As aforementioned, we usually take the $z$-axis as the quantization axis and the $ x$ and $ y$-axes as the driving axes in the context of NMR. We, however, keep our notations above: the $x$-axis is the quantization axis while the $ y$ and $ z$-axes are driving axes. If one wants to reproduce expressions in the NMR notations, it is sufficient to replace $x\rightarrow z$, $z\rightarrow y$, and $y\rightarrow x$.

\subsection{Short-CORPSE and Improvement}
Let $R_{z}(\theta)$ denote the rotation with angle $\theta$ and the direction $z$ in the Bloch sphere representation, i.e.,
\begin{align}
	R_{z}(\theta)&=\cos(\theta/2){\mathbb I}_{2}-i\sin(\theta/2)\sigma_{z}=
	\begin{pmatrix}
		e^{-i\theta/2}&0\\
		0&e^{i\theta/2}
	\end{pmatrix}
	.
	\label{eq:target}
\end{align}
Our goal is to implement $R_{z}(\theta)$ as the zeroth-order operation while keeping error robustness. In what follows, when a pulse implements $R_{z}(\theta)$ as the zeroth-order operation, we call it a $\theta$-pulse.
Comparing Eq.~\eqref{eq:parameterize} and \eqref{eq:target}, one find
\begin{equation}
	\theta=\int^{T}_{0}\Omega(t) dt. 
\end{equation}

The pulse shape of Short-CORPSE \cite{Cummins_2000,PhysRevA.67.042308} to realize $R_{z}(\theta)$ is given by
\begin{align}	
	\Omega(t) &=	\begin{cases}		
		-1 & \qquad 0 \le t < \theta_{\rm 1} \\		
		1 & \qquad  \theta_1 \le t < \theta_{\rm 1}+ \theta_{\rm 2}\\		
		-1 &  \qquad   \theta_{\rm 1}+ \theta_{\rm 2} \le t < 2\theta_{\rm 1}+ \theta_{\rm 2}	\end{cases},
\end{align}
where $\displaystyle \theta_1 = \pi - \kappa - \theta/2, \theta_2 = 2\pi - 2\kappa $ and $\displaystyle \kappa = \sin^{-1} \left( \sin (\theta/2)/2 \right)$. The corresponding first-order error $g_{1}(t)$ is obtained by integrating  Eq.~\eqref{rec_rel_1} with the initial condition of $g_1(0) =0$.
\begin{align}
	g_{1}(t)&=\begin{cases}
		-i\left(e^{i t }-1\right) &0 \le t < \theta_1 \\
		i\left(1+e^{-i(t-2\theta_1)}-2e^{2i\theta_1}\right) & \theta_1 \le t < \theta_1+ \theta_2\\
		i\left(1-e^{i(t-2\theta_2)}+2\left(e^{i(\theta_1-\theta_2)}-e^{i\theta_1}\right)\right)   
		& \theta_1+ \theta_2 \le t < 2\theta_1+ \theta_2
		\end{cases}.
\end{align}
The error trajectory is obtained by $(x(t),y(t))=(\Re{(g_{1}(t))}, \Im{(g_{1}(t))})$.We define the area enclosed by the trajectory as $S_{\theta}$, whose explicit form is given as
\begin{align}
	S_{\theta}&=\frac{\theta +\sin \theta +\sqrt{14+2\cos\theta}\sin \left(\theta/2\right)}{2}.
\end{align} 
To implement the second-order error robustness, or equivalently, the zero net area condition while keeping the trajectory closed, we cancel the area $S_{\theta}$ by adding an opposite-directed circle trajectory with radius $r=\sqrt{S_{\theta}/\pi}$: This is achieved by the $2\pi$-pulse with $\Omega(t)=1/r$ during the time interval of $2\pi r$. The pulse shape of this new DCQG is given by
\begin{align}		
	\label{n_p_shape}	
	\Omega(t) &=		\begin{cases}				
		-1 & \qquad 0 \le t < \theta_{\rm 1} \\				
		1 & \qquad  \theta_1 \le t < \theta_{\rm 1}+ \theta_{\rm 2}\\				
		-1 &  \qquad   \theta_{\rm 1}+ \theta_{\rm 2} \le t < 2\theta_{\rm 1}+ \theta_{\rm 2}	\\		
		-1/r &  \qquad   2\theta_{\rm 1}+ \theta_{\rm 2} \le t < 2\theta_{\rm 1}+ \theta_{\rm 2} + 2\pi r	
	\end{cases}.
\end{align}
Note that $2\pi$-pulse does not affect the first-order robustness because the start and end of its error trajectory are the same point.

\subsection{Short-CORPSE of  $3\pi/2$ rotation}

\begin{figure}[h] 
	\begin{center}
		\includegraphics[width=60mm]{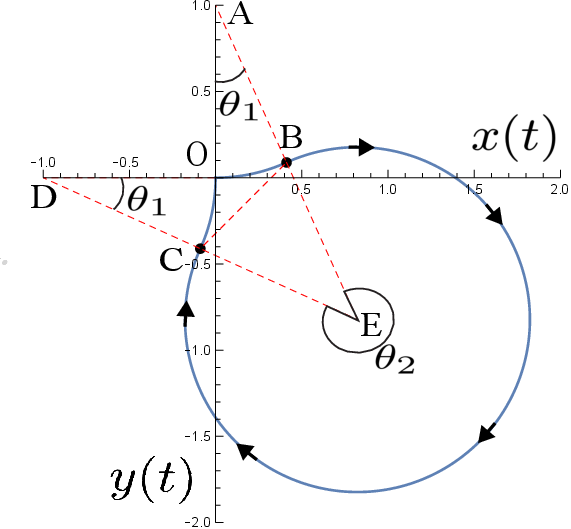}
		\caption{The error trajectory of the Short-CORPSE of $3\pi/2$ rotation: O $\rightarrow$ B $(\frac{-1+\sqrt{7}}{4}, \frac{3-\sqrt{7}}{4})$ in $t \in [0, \theta_{1})$,  B $\rightarrow$ C $(\frac{-3+\sqrt{7}}{4}, \frac{1-\sqrt{7}}{4} )$ in $t \in [\theta_{1}, \theta_{1}+\theta_{2})$, and C $\rightarrow$ O in $t \in [ \theta_{1}+\theta_{2}, 2 \theta_{1}+\theta_{2}]$, respectively. Note that E $(\frac{-1+\sqrt{7}}{2}, \frac{1-\sqrt{7}}{2} )$.
		\label{Area_cal}}
	\end{center}
\end{figure}

One qubit rotation of the angle $\theta = \pi/2$ is important in controlling a quantum system, and its equivalent operation is a $3\pi/2$ rotation. The Short-CORPSE of  $3\pi/2$ rotation is shorter than that of  $\pi/2$ \cite{Kukita_2024} and thus we first consider it. Fig.~\ref{Area_cal} shows the error trajectory $(x(t), y(t))$ of the Short-CORPSE of  $3\pi/2$ rotation. 

\begin{figure}[h]
	\begin{center}
		\includegraphics[width=80mm]{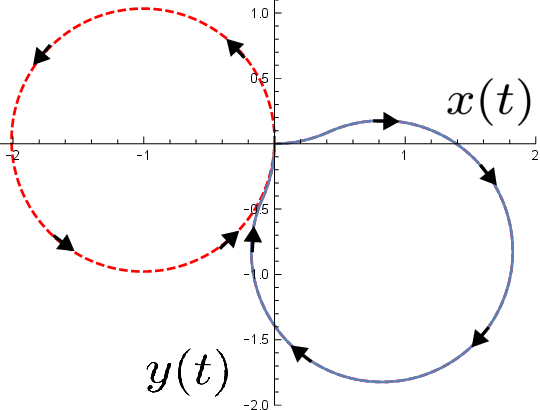}
		\caption{The error trajectory of the Short-CORPSE and $2\pi$ rotation: The blue solid curve (clockwise) corresponds to that of the Short-CORPSE of $\displaystyle 3\pi/2$ rotation, while the red-dashed (counterclockwise) corresponds to that of the added $2\pi$ pulse.  The strength of this pulse is $1/r$. 
		\label{2ndOEC}}
	\end{center}
\end{figure}

The area $S_{3\pi/2}$ enclosed by the error trajectory for $R_{z}(3\pi/2)$ is calculated with Fig.~\ref{Area_cal}.
\begin{align}
	S_{3\pi/2}&= S({\rm sector\, EBCE})+ S({\rm triangle\, ECB})+S({\rm polygon\, OABCDO} )
	\nonumber \\
	&- S({\rm sector\, ABOA})-S({\rm sector\,  DCOD})
	\nonumber \\
	&= \frac{2 \sqrt{7} -2 + 3\pi}{4} \sim 1.01193 \pi, 
\end{align}
where $S(*)$ denotes the area of a figure $*$. Then we obtain $\displaystyle r= \sqrt{S_{3\pi/2}/\pi}$ in Eq.~\eqref{n_p_shape}. Figure~\ref{2ndOEC} shows the error trajectory of the Short-CORPSE and this $2\pi$ rotation. 

\begin{figure}[h]
	\begin{center}
		\includegraphics[width=80mm]{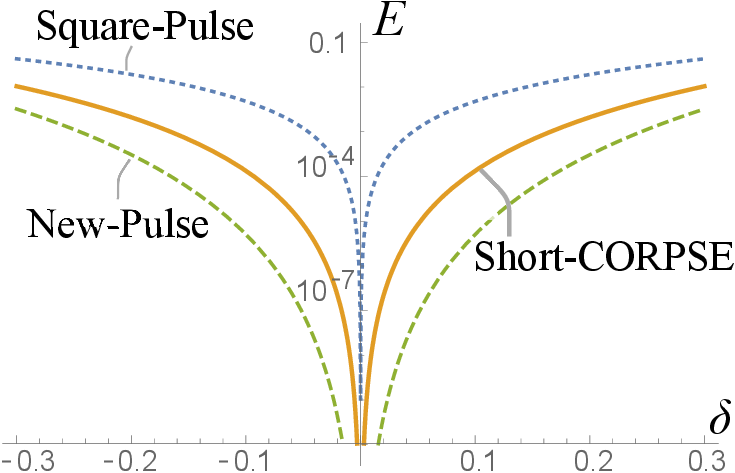}
		\caption{Infidelity of three $3\pi/2$ rotations under ORE whose signed strength is $\delta$.  Square-Pulse (without ORE correction), Short-CORPSE (with the first-order ORE correction), and the new pulse sequence designed to correct ORE up to the second-order based on geometric considerations.  
		\label{inf_h_pi}}
	\end{center}
\end{figure}

To consider the performance of the pulse sequence, we introduce the infidelity:
\begin{align}
E&=1-\frac{{\rm Tr}\left(R^{\dagger}_{z}(\theta)U_{\delta}(T)\right)}{2},
\end{align}
where $R_{z}(\theta)$ is the target operation while $U_{\delta }(T)$ is the actual one caused by the driving field $\Omega(t)$ subject to ORE. Figure~\ref{inf_h_pi} compares the infidelity of three pulse sequences ($3\pi/2$ rotation) when ORE exists.  Square-Pulse (without ORE correction), the Short-CORPSE (with the first-order ORE correction), and the new pulse sequence designed to correct ORE up to the second-order based on geometric considerations. 

The strength $1/r \sim 0.994$ is very close to 1. Thus, we can take $1/r = 1$ when a slightly imperfect cancellation is allowed: This will be useful, as one can maintain the same pulse strength and adjust only the phases of the individual pulses.

\subsection{Short-CORPSE of $\pi$ rotation}

One qubit rotation of the angle $\theta = \pi$ is also important in controlling a quantum system. It is sometimes called a population inversion gate. We calculate the area $S_{\pi}$ enclosed by the error trajectory for the Short-CORPSE of $\pi$ rotation, similarly to in the case of the $3\pi/2$ rotation, and obtain $\displaystyle S_{\pi} = \frac{1}{2}\left(2\sqrt{3}+\pi\right) \sim 1.051 \pi $. 

\begin{figure}[h]
	\hspace{0mm} (a) \hspace{55mm} (b) \vspace{-10mm} 
	\begin{center}
		\includegraphics[width=55mm]{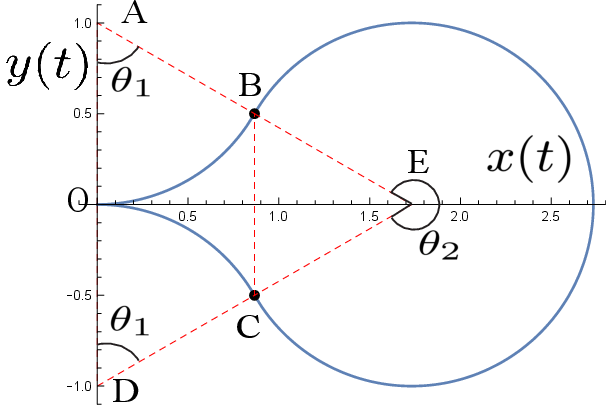} \hspace{15mm}
		\includegraphics[width=55mm]{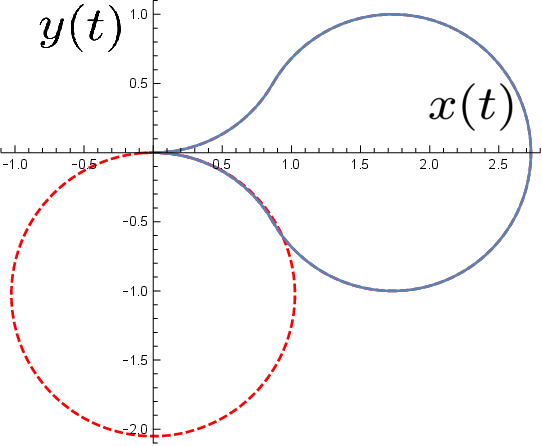} 
		\caption{Error trajectories. (a) Short-CORPSE of $\pi$ rotation: O $\rightarrow$ B $(\frac{\sqrt{3}}{2}, \frac{1}{2})$ in $t \in [0, \frac{\pi}{3} )$,  B $\rightarrow$ C $(\frac{\sqrt{3}}{2}, \frac{1}{2})$ in $t \in [\frac{\pi}{3}, \frac{6\pi}{3})$, and C $\rightarrow$ O in $t \in [ \frac{6\pi}{3}, \frac{7\pi}{3}]$, respectively. Note that E $(\sqrt{3}, 0 )$ and $\theta_1 = \frac{\pi}{3}, \theta_2 = \frac{5\pi}{3}$. (b) Short-CORPSE and $2\pi$ rotation: The blue solid curve (clockwise) corresponds to the error trajectory of the Short-CORPSE of $\displaystyle \pi$ rotation, while the red-dashed (counterclockwise) corresponds to that of the added $2\pi$ pulse.  
			\label{2ndOECPi}}
	\end{center}
\end{figure}	

\begin{figure}[h]
	\begin{center}
		\includegraphics[width=80mm]{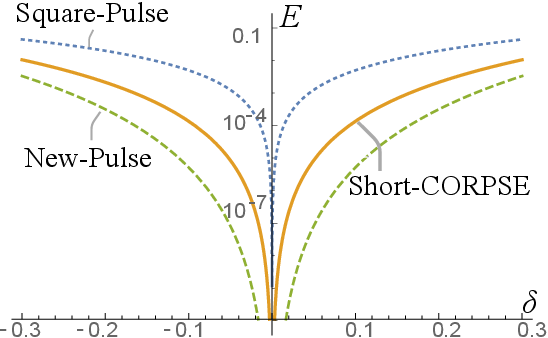}
		\caption{Infidelity of three $\pi$ rotations under ORE whose signed strength is $\delta$.  Square-Pulse (without ORE correction), Short-CORPSE (with the first-order ORE correction), and the new pulse sequence designed to correct ORE up to the second-order based on geometric considerations.  
		\label{inf_pi}}
	\end{center}
\end{figure}

Figure~\ref{inf_pi} compares the infidelity of three pulse sequences ($\pi$ rotation): Square-Pulse (without ORE correction), the Short-CORPSE (with the first-order ORE correction), and the new pulse sequence designed to correct ORE up to the second-order based on geometric considerations. 

\section{Conclusion}\label{conc}

Off resonance error (ORE) is one of the most typical errors in one-qubit control. Implementation of dynamically corrected quantum gates (DCQG) robust against this error will be important for realization of quantum technologies. Many ORE-robust DCQGs that compensate for the first order of the error are known; on the other hand, second-order ORE-robust ones have less been studied. We propose a method to promote ORE-robust DCQG up to the first order of the error to that up to the second order using a geometric aspect. We use Short-CORPSE, a famous first-order ORE-robust DCQG, as a seed, and turn it to a second-order DCQG. It is done by only adding a $2\pi$-pulse: a minimum modification can improve the performance of known pulse sequences. Our technique is applicable to other first-order DCQGs. The NMR community can apply the insights we have gained so far to their measurements \cite{SR_Comp_Pulse_2020}, and these insights are also applicable to other quantum systems \cite{JONES202449}.

\bmhead{Acknowledgements}

 This work was supported by JSPS Grants-in-Aid for Scientific Research (25K07194).


\bibliography{25Aug18}

\end{document}